\documentclass[12pt]{article}
\usepackage[english]{babel}
\usepackage[utf8]{inputenc}
\usepackage[T1]{fontenc}
\def\tGamma{\tilde{\Gamma}}
\def\mF{\mathcal{F}}
\usepackage{amsmath}
\usepackage{amsfonts}
\usepackage{url}
\usepackage[dvips]{graphicx}
\usepackage{calc}
\usepackage{subfigure}
\usepackage{multirow}

\def\mL{\mathcal{L}}
\def\mH{\mathcal{H}}

\def\bA{\mathbf{A}}

\def\mG{\mathcal{G}}
\newcommand{\tg}{\tilde{g}}

\begin{document}
	\begin{center}
		{\Large{ \bf Weyl Gravity in Covariant Hamiltonian Formalism}}
		
		\vspace{1em}  J. Kluso\v{n}\textsuperscript{\textdagger} and B. Matou\v{s}\textsuperscript{\textdagger}\textsuperscript{\textdaggerdbl}
		\footnote{Email addresses:
			J. Kluso\v{n}:\ klu@physics.muni.cz, B. Matou\v{s}:\  bmatous@mail.muni.cz}\\
		\vspace{1em} \textsuperscript{\textdagger} \textit{Department of Theoretical Physics and
			Astrophysics, Faculty
			of Science,\\
			Masaryk University, Kotl\'a\v{r}sk\'a 2, 611 37, Brno, Czech Republic}\\
		\vspace{1em} \textsuperscript{\textdaggerdbl} \textit{North-Bohemian Observatory and Planetarium in Teplice, \\
			Kopern\'{i}kova 3062, 415 01, Teplice,
			Czech Republic}\\
		
	\end{center}
	
	\abstract{We find covariant canonical formalism for Weyl invariant gravity. We discuss constraint structure of this theory and its gauge fixed form.}

\section{Introduction and Summary}\label{first}
It is well known that theories with reduced  diffeomorphism invariance are far less restricted than  diffeomorphism invariant theories, striking example is famous Ho\v{r}ava-Lifschitz gravity
\cite{Horava:2009uw,Horava:2010zj}. Another example of theories with restricted diffeomorphism are theories invariant under  transverse diffeomorphisms and Weyl transformations \cite{Alvarez:2006uu,Alvarez:2012px,Oda:2016psn}. These theories offer very interesting alternative to General Relativity (GR) and they firstly emerged with the observation that theory of self-interacting gravitons does not need to be General Relativity. Instead such alternatives could be Weyl transverse gravities (WTG) or unimodular gravities, for recent 
review, see for example 
\cite{Alvarez:2023utn,Jirousek:2023gzr,Carballo-Rubio:2022ofy}. It can be shown that classical solutions of WTG and GR equations of motions are equivalent however WTG or their gauge fixed version which is unimodular gravity imply that cosmological constant is radiative 
stable \cite{Carballo-Rubio:2015kaa}, for recent extended discussion see 
\cite{Jirousek:2023gzr}. Another interesting check of the consistency of WTG was given in \cite{Alonso-Serrano:2022rzj,Alonso-Serrano:2022pif}, where Noether charge formalism for these theories was developed. We would like again stress that this is non-trivial result  due to the restricted diffeomorphism invariance of WTG theories. 

Since WTG theory possesses many interesting properties we mean that it is natural to study WTG from different point of view. In this paper we focus on covariant canonical formulation, known as 
 Weyl-De Donder formalism
\cite{DeDonder,Weyl}, 
of this theory. Main advantage of Weyl-De Donder formalism is 
that it treats all partial derivatives as equivalent  when we define conjugate momenta which is especially useful in case of manifestly diffeomorphism invariant theories.  
This alternative treatment of the canonical formalism of field theories was further developed for example in 
\cite{Struckmeier:2008zz,Kanatchikov:1997wp,Forger:2002ak}, for review, see \cite{Kastrup:1982qq}\footnote{For another interesting applications of covariant canonical formalism, see for example \cite{Lindstrom:2020szt,Kluson:2020pmi}.}.

In order to find covariant canonical formalism of WTG theory we should proceed in similar way as in case of Einstein-Hilbert action \cite{Parattu:2013gwa,Horava:1990ba}
when we  split Lagrangian into bulk part and boundary part.
In case of WTG theory we should be very careful due to absence  of the determinant 
of metric in the action and we find that corresponding form of bulk Lagrangian is different from 
Einstein-Hilbert action. Then we proceed to the definition of conjugate momenta. Following very careful analysis presented in \cite{Parattu:2013gwa,Padmanabhan:2013nxa} we introduce new variable $f^{ab}$ instead of $g^{ab}$ that is related to $g^{ab}$ by  point transformation $f^{ab}=\sqrt{-g}g^{ab}$. An importance of this variable was already stressed in \cite{Edington,Schrodinger,Einstein:1955ez}. As was argued in \cite{Parattu:2013gwa} canonical form of  Einstein Hilbert action has remarkable simple form expressed with the help of variables  $(f^{ab},N^c_{ab})$ and it is also independent on square root of $f$. In case of WTG gravity the situation is slightly different when introducing $f^{ab}$ and conjugate momentum $N_{ab}^c$ again simplifies canonical form of the action significantly however the Hamiltonian still depends on the polynomial of the determinant of matrix $f^{ab}$. On the other hand we will show that this fact is crucial for the preservation of the primary constraints $\mG^c\equiv f^{ab}N_{ab}^c$ whose presence is reflection of the invariance of the action under Weyl rescaling of metric. In fact, in terminology of Dirac constrains system it is natural to call $\mG^c$ as the first class constraint. Then we show that this gauge symmetry can be naturally fixed by introducing unimodular constraint 
$\sqrt{-f}=K$ where $K$ is constant. In other words we reproduce using covariant canonical formalism that gauge fixed version of WTG is unimodular gravity. Again this is rather non-trivial result due to the fact that it is not completely clear how to deal with constraint systems in covariant canonical gravity.

As the next step we perform covariant canonical analysis of Weyl gravity which is formulated without auxiliary metric
\footnote{For recent discussion of this theory, see for example
\cite{Alvarez:2023zqz}.}. We again perform   splitting of the Lagrangian into bulk and boundary term. Then we introduce new variable $f^{ab}=(-g)^\alpha g^{ab}$ where $\alpha$ is arbitrary parameter. We choose general $\alpha$ in order to  analyze possible dependence of the Hamiltonian on $\alpha$. Surprisingly we find that the Hamiltonian does not depend on $\alpha$ at all. This is very remarkable result. Then we identify corresponding Hamiltonian and primary constraints and we show that they have exactly the same form as in case of the WTG theory formulated in terms of auxiliary metric.  Finally we express the boundary Lagrangian as function of canonical variables and we show that it can be derived from the bulk part of the Lagrangian which is in agreement with the holographic relation between bulk and boundary Lagrangians as was shown for example in \cite{Mukhopadhyay:2006vu}. We mean that this is again non-trivial result due to the fact that WTG theory is not invariant under full diffeomorphism. 

Let us outline our results and suggest possible extension of this work. We found covariant canonical formalism for WTG gravity. We identified primary constraint which is generator of Weyl transformation. We also found corresponding equations of motion and we argued that  this gauge freedom can be fixed by unimodular constraint. On the other hand there is an important problem in this analysis which is the fact that the equations of motion of gauge fixed WTG gravity do not reproduce equations of motion of unimodular gravity that were derived recently in \cite{Kluson:2023yzv}.  Unfortunately we are not able to identify origin of non-equivalence of these two formulations. It is possible that they are hidden in the basic structure of covariant canonical formalism or our approach how we deal with the constraints in covariant canonical gravity is too naive and more powerful techniques, as for example 
one developed by Kanatchikov in \cite{Kanatchikov:1997wp} could be more appropriate for this analysis. We hope to return to this problem in future.  We also found covariant Hamiltonian for WTG theory formulated without auxiliary metric and we determined the boundary term as function of canonical variables. We also shown that this boundary term 
can be expressed with the variation of the bulk term with respect to the derivative of canonical variable which is in agreement with the holographic interpretation of WTG gravity. 

This paper is organized as follows. In the next section (\ref{second}) we introduce WTG gravity formulated with the auxiliary metric and we determine corresponding covariant Hamiltonian. Then in section (\ref{third}) we perform the same analysis in case of WTG gravity formulated in terms of physical metric and we again find corresponding Hamiltonian and primary constraints.

\section{Weyl Invariant Theory of Gravity in Covariant Formalism}\label{second}
In this section we present basic facts about  Weyl invariant gravity and we find its covariant form.  The natural formulation of  Weyl invariant gravity is based on an introduction of  auxiliary metric 
\begin{equation}\label{auxmetric}
\tg_{ab}=\left(\frac{\omega^2}{-\det g}\right)^{\frac{1}{n}}g_{ab} \ 
\end{equation}
that is manifestly invariant under rescaling 
\begin{equation}
g'_{ab}(x)=\Omega (x)g_{ab}(x) \ . 
\end{equation}
Note that $n$ labels number of space-time dimensions. 
Further, $\omega(x)$ can be generally $n$ dimensional volume form. For simplicity we will presume that $\omega$ is constant.  Then we can write an action for Weyl gravity in the form \cite{Alvarez:2012px,Oda:2016psn}
\begin{equation}
S=\int d^{n}x\mL \ , \quad 
\mL=\frac{1}{16\pi}\omega \tilde{R}(\tilde{g}) \ . 
\end{equation}
In order to find covariant Hamiltonian formulation of 
Weyl gravity it is natural to split 
 Lagrangian into bulk and boundary term. Recall that $\tilde{R}$ can be written as
\begin{eqnarray}
&&\tilde{R}=\tilde{Q}_k^{ \ mnl}\tilde{R}^k_{ \ mnl} \ , \nonumber \\
&&\tilde{R}^k_{ \ mnl}=\partial_n \tilde{\Gamma}^k_{lm}-
\partial_l \tilde{\Gamma}^k_{nm}+\tilde{\Gamma}^k_{np}\tilde{\Gamma}^p_{lm}
-\tilde{\Gamma}^k_{lp}\tilde{\Gamma}^p_{mn} \ , \nonumber \\
&&\tilde{Q}_k^{ \ mnl}=\frac{1}{2}(\tg^{ml}\delta_k^n-\tg^{mn}\delta_k^l)
\ . \nonumber \\
\end{eqnarray}
From the definition of $\tilde{Q}$ we get that it is anti-symmetric in the two last indices $\tilde{Q}_k^{mnl} = -\tilde{Q}_k^{mln}$.
%
%
Then we can write the scalar curvature as 
\begin{equation}
\tilde{R} = 
2\partial_n(\tilde{Q}_k^{ \ mnl} \tGamma_{lm}^k)
-2\tGamma_{lm}^k \partial_n\tilde{Q}_k^{ \ mnl}
+2\tilde{Q}_k^{ \ mnl}\tGamma_{np}^k\tGamma^p_{lm}
\:,
\end{equation}
from this we immediately get both parts of Lagrangians. The boundary part
\begin{equation}
    \mL_{bound} = 
    \frac{\omega}{16\pi} \partial_n(2\tilde{Q}_k^{ \ mnl} \tGamma_{lm}^k) =
    \frac{\omega}{16\pi} \partial_n(\tilde{g}^{ml} \tGamma_{lm}^n - \tilde{g}^{mn} \tGamma_{lm}^l)\:,
\end{equation}
and the bulk part
\begin{eqnarray}
&&    \mL_{bulk} = 
    \frac{\omega}{8\pi}\tilde{Q}_k^{ \ mnl}\tGamma_{np}^k\tGamma^p_{lm} - \frac{\omega}{8\pi}\tGamma_{lm}^k \partial_n\tilde{Q}_k^{ \ mnl} =\nonumber \\
&&=\frac{\omega}{16\pi}  \left( \tilde{g}^{mn} \tGamma^l_{np} \tGamma^p_{lm} - \tilde{g}^{mn} \tGamma^l_{ml} \tGamma^p_{np} \right)
    \:, \nonumber \\
\end{eqnarray}
where we have used
\begin{equation}
    2\partial_n \tilde{Q}_k^{ \ mnl} =
    \delta^l_k (\tGamma^m_{np} \tilde{g}^{pn} + \tGamma^n_{np} \tilde{g}^{mp}) 
    - \tGamma^m_{kp} \tilde{g}^{pl} - \tGamma^l_{kp} \tilde{g}^{mp} \:.
\end{equation}
%
%
%
%
%
%
Before we proceed to the covariant canonical formalism we should stress one important point which is the fact that $\tGamma^r_{ra}$ vanishes identically.
In more details,  writing $\tg_{mn}$ as $\tg_{mn}=\Omega g_{mn} \ , \Omega=\frac{\omega^{\frac{2}{n}}}{(-g)^{\frac{1}{n}}}$ we get 
\begin{eqnarray}
\tGamma^r_{ri}=\frac{1}{2}\tg^{rm}\partial_i \tg_{mr}=\frac{1}{2}\frac{\partial_i g}{g}
+\frac{n}{2\Omega}\partial_i \Omega=0
\end{eqnarray}
as follows from the fact that 
\begin{equation}
\partial_i \Omega=-\frac{\Omega}{n}\frac{\partial_i g}{g} \ . 
\end{equation}
Then using  the condition $\tGamma^r_{ri}=0$ the Lagrangian simplifies considerably
\begin{eqnarray}\label{mLdef}
&&\mL=\mL_{bound}+\mL_{bulk} \ , \nonumber \\
&&\mL_{bulk}=\frac{\omega}{16\pi}\tGamma^m_{nk}\tg^{kl}\tGamma^n_{lm} \ , \quad 
\mL_{bound}= \frac{\omega}{16\pi}\partial_n\left[\tg^{ml}\tGamma^n_{lm}\right] \ . \nonumber \\
\end{eqnarray}
Now we are ready to find   covariant canonical formulation of WTG gravity.  As the first step we introduce  suitable canonical variables. Recall that the theory is formulated with the help of auxiliary metric (\ref{auxmetric}).
At first sight we should select $g_{mn}$ as the canonical variable. On
 the other hand it was argued by Padmanabhan in many places, see for example 
\cite{Parattu:2013gwa}, that natural variable for the study of dynamics of 
gravity should be chosen $f^{ab}$ that is defined as
\begin{equation}\label{deffab}
f^{ab}=\sqrt{-g}g^{ab} \ .
\end{equation} In fact, an importance of this object 
was already stressed in 
\cite{Einstein:1955ez,Edington,Schrodinger}. Then it is natural to find direct relation between $\tg_{mn}$ and $f^{mn}$. First of all from  (\ref{deffab}) we obtain 
\begin{equation}
    f=\det f^{ab} ,\quad (-f)=(-g)^{\frac{n-2}{2}} ,\quad   (-g)=(-f)^{\frac{2}{n-2}} \:
\end{equation}
Then after some manipulation we get direct relation between $\tg_{ab}$ and $f^{ab}$ in the form 
\begin{equation}\label{pointtr}
\tg^{mn}
=\left(\frac{1}{-\omega^2 f}\right)^{\frac{1}{n}}f^{mn} ,\quad \det \tg^{mn}=-\frac{1}{\omega^2}   \ ,
\end{equation}
where $\tg^{mn}$ is inverse to $\tg_{mn} \  ,\tg_{mn}\tg^{nk}=\delta_m^k$. Clearly (\ref{pointtr}) is point transformation.

Having selected $f^{ab}$ as canonical variable we are ready to determine corresponding conjugate momenta as  
\begin{eqnarray}
&&N^c_{ab}=\frac{\partial\mL_{bulk}}{\partial (\partial_c f^{ab})}=
\frac{\partial \mL_{bulk}}{\partial (\partial_k \tg_{mn})}
\frac{\partial (\partial_k \tg_{mn})}{\partial(\partial_c f^{ab})} =
\nonumber \\
&&=-M^{kmn}
\left(\tg_{mr}\frac{\partial (\partial_k\tg^{rs})}{\partial (\partial_c f^{ab})}\tg_{sn}\right)\ , \nonumber \\
\end{eqnarray}
where $M^{kmn}$ is defined as
\begin{eqnarray}
&&M^{kmn}=    \frac{\partial \mL_{bulk}}{\partial(\partial_k \tg_{mn})}=
\frac{\omega}{8\pi}
\frac{\partial \tGamma^r_{ps}}{\partial(\partial_k\tg_{mn})}
\tg^{sl}\tGamma^p_{lr}
=\frac{\omega}{16\pi}\tg^{mr}\tGamma_{rl}^k\tg^{ln} \ , 
\nonumber \\
\end{eqnarray}
as follows from definition of $\mL_{bulk}$  given in (\ref{mLdef}) and
where we also used following variation 
\begin{eqnarray}
\frac{\partial \tGamma^r_{ps}}{\partial (\partial_k\tg_{mn})}
=\frac{1}{4}\tg^{rt}(\delta_p^k(\delta_t^m\delta_s^n+\delta_t^n\delta_s^m)+
\delta_s^k(\delta_t^m\delta_p^n+\delta_t^n\delta_p^m)
-\delta_t^k(\delta_p^m\delta_s^n+\delta_p^n\delta_s^m) \ . 
\nonumber \\
\end{eqnarray}
Since (\ref{pointtr}) is point transformation we obtain
\begin{eqnarray}
&&    \frac{\partial (\partial_k \tg^{rs})}{\partial (\partial_c f^{ab})}=
    \delta_k^c\frac{\partial \tg^{rs}}{\partial f^{ab}}=\nonumber \\
&&    =\delta_k^c
\frac{1}{\omega^\frac{2}{n}}(-f)^{-\frac{1}{n}}
\left[\frac{1}{2}
(\delta_a^r\delta_b^s+\delta_a^s\delta_b^r)-\frac{1}{n}f_{ab}f^{rs}\right] \equiv \delta_k^c B^{rs}_{ab} \: .
\nonumber \\
\end{eqnarray}   
Then the momentum conjugate to $f^{ab}$ is equal to
\begin{eqnarray}\label{Nab}
&&	N^c_{ab}
=
	-\frac{\omega}{16\pi}\tGamma^c_{rs}
 \frac{1}{\omega^\frac{2}{n}}(-f)^{-\frac{1}{n}}
	[\frac{1}{2}
	(\delta_a^r\delta_b^s+\delta_a^s\delta_b^r)-\frac{1}{n}f_{ab}f^{rs}]=
	\nonumber \\
&&	=-\frac{\omega^\frac{n-2}{n}}{16\pi }(-f)^{-\frac{1}{n}}
	[\tGamma^c_{ab}-\frac{1}{n}\tGamma^c_{rs}f^{rs}f_{ab}] \ . 
	\nonumber \\
\end{eqnarray}
From (\ref{Nab}) we immediately obtain $N^c_{ab}f^{ab}=0$ and hence 
we have $n$ primary constraints
\begin{equation}
\mG^c\equiv N^c_{ab}f^{ab}\approx 0 \ . 
\end{equation}
As the next step we determine  bare Hamiltonian that is defined as 
\begin{equation}
\mH_B=\partial_c f^{ab}N_{ab}^c-\mL_{bulk} \ . 
\end{equation}
Since $\mL_{bulk}$ is function of $\tg_{mn}$ instead of $f^{ab}$
it is natural to perform following manipulation
\begin{eqnarray}
	\partial_c f^{ab}N_{ab}^c=
-	\partial_c f^{ab}M^{kmn}\tg_{mr}\delta_k^c
	B^{rs}_{ab}\tg_{sn}=\partial_k \tg_{mn}M^{kmn}
\end{eqnarray}
using the fact that
\begin{eqnarray}
	\partial_k \tg_{mn}=-\tg_{mr}
	\partial_k \tg^{rs}\tg_{sn}=
	-\tg_{mr}\frac{\delta \tg^{rs}}{\delta f^{ab}}\partial_k f^{ab}\tg_{sn}=-\tg_{mr}B^{rs}_{ab}\partial_k f^{ab}\tg_{sn} \ . \nonumber \\
\end{eqnarray}
Then the Hamiltonian density has the form 
\begin{eqnarray}
	\mH_B=\partial_c f^{ab}N_{ab}^c-\mL_{bulk}=\partial_k\tg_{mn}M^{kmn}-
	\mL_{bulk}
	=\frac{\omega}{16\pi}\tGamma^a_{cm}\tg^{mb}\tGamma^c_{ab} \ . 	\nonumber \\
\end{eqnarray}
Finally we should express Hamiltonian density as function of canonical variables. This is slightly problematic due to the fact that the relation between $\tGamma^a_{ab}$ and $N^c_{ab}$ is not invertible. For that reason let us calculate following combination
\begin{eqnarray}
	N^c_{ab}\tg^{ad}N^b_{dc}=
	\bA^2[\tGamma^c_{ab}\tg^{ad}\tGamma^b_{dc}
	+\frac{1}{n^2}\tGamma^c_{rs}\tg^{rs}\tg_{cb}
	\tGamma^b_{mn}\tg^{mn}] \ , \nonumber \\
\end{eqnarray}
where $\bA= -\frac{\omega^\frac{n - 2}{n}}{16\pi }(-f)^{-\frac{1}{n}}$. We further
have 
\begin{eqnarray}
	N^r_{ra}\tg^{ab}N^t_{tb}=
	\frac{\bA^2}{n^2}\tGamma^a_{rs}\tg^{rs}\tg_{ab}\tGamma^b_{mn}
	\tg^{mn}\ .  \nonumber \\
\end{eqnarray}
Collecting these terms together we obtain that the 
 Hamiltonian density is equal to
\begin{eqnarray}
	\mH_B
	=\frac{16\pi}{\omega^{\frac{n-2}{n}}}(-f)^{\frac{1}{n}}
	[   N^c_{ab}f^{ad}N^b_{dc}
	-N^r_{ra}f^{ab}N^t_{tb}]\ . \nonumber \\
\end{eqnarray}
Then the canonical form of the action has the form 
\begin{eqnarray}
	S=\int d^nx (\partial_c f^{ab}N^c_{ab}-\mH_B-\Lambda_c\mG^c) \ ,  \nonumber \\
\end{eqnarray}
where we included primary contraints $\mG^c\approx 0$ multiplied by Lagrange multipliers 
$\Lambda_c$. Note that we treat $\Lambda_c$ as independent variables which should be varied when we search for extrema of the action. Explicitly, the variation of the action has the form 
\begin{eqnarray}
&&	\delta S=\int d^nx\left(\partial_c \delta f^{ab}N^c_{ab}+
	\partial_c f^{ab}\delta N^c_{ab}-\frac{\delta \mH_B}{\delta f^{ab}}\delta f^{ab}-\frac{\delta \mH_B}
	{\delta N^c_{ab}}\delta N^c_{ab}-\right.\nonumber \\
&& \left.	-\delta \Lambda_c \mG^c
	-\Lambda_d\frac{\delta \mG^d}{\delta f^{ab}}\delta f^{ab}
	-\Lambda_d \frac{\delta \mG^d}{\delta N^c_{ab}}\delta N^c_{ab}\right)=0 \nonumber \\
\end{eqnarray}
that gives following equations of motion 
\begin{eqnarray}
&&	\partial_c f^{ab}-\frac{\delta \mH_B}{\delta N^c_{ab}}-\Lambda_d\frac{\delta \mG^d}{\delta N^c_{ab}}=0 \ , 
	\nonumber \\
&&	\partial_c N^c_{ab}+\frac{\delta \mH_B}{\delta f^{ab}}+
	\Lambda_d\frac{\delta \mG^d}{\delta f^{ab}}=0 \ , 
	\nonumber \\
&&	\mG^c=0 \ , \nonumber \\
\end{eqnarray}
or explicitly
\begin{eqnarray}\label{eqmo}
	&&    \partial_c f^{ab}     -\Lambda_c f^{ab}  -\nonumber \\
	&&    -\frac{16\pi}{\omega^{\frac{n-2}{n}}}(-f)^{\frac{1}{n}}  
	\left(
	f^{db} N^a_{cd} + f^{da} N^b_{cd} - \delta^a_c f^{bs} N^m_{ms} - \delta^b_c f^{as} N^m_{ms}
	\right)=0 \ , \nonumber \\
	&&     \partial_c N^c_{ab} + \Lambda_c N^c_{ab} + \nonumber \\
	&&    + \frac{16\pi}{n \omega^{\frac{n-2}{n}}} (-f)^{\frac{1}{n}} f_{ab} (N^m_{cd}f^{dn}N^c_{mn}-N^m_{ms}f^{sr}N^n_{nr})  +  \nonumber \\
	&&   +\frac{16\pi}{\omega^{\frac{n-2}{n}}}(-f)^{\frac{1}{n}} (N^m_{na} N^n_{mb} - N^m_{ma} N^n_{nb})=0 
	\ , \quad N^c_{ab}f^{ab}=0 \ . \nonumber \\
\end{eqnarray}
Let us now return to the constraint $\mG^c\approx 0$ and study its time evolution. From the equations of motion above we get
\begin{equation}\label{contfabNc}
	N^c_{ab} \partial_c f^{ab} =
	\Lambda_c N^c_{ab} f^{ab} +
	\frac{32\pi}{\omega^{\frac{n-2}{n}}}(-f)^{\frac{1}{n}}  
	\left(
	f^{db} N^a_{cd} N^c_{ab} - N^c_{cb} f^{bs} N^m_{ms}
	\right) \:,
\end{equation}
\begin{eqnarray}
&&	f^{ab} \partial_c N^c_{ab} =
	- \Lambda_c N^c_{ab} f^{ab}
	- \frac{16\pi}{\omega^{\frac{n-2}{n}}}(-f)^{\frac{1}{n}}  
	\left( f^{db} N^a_{cd} N^c_{ab}- N^c_{cb} f^{bs} N^m_{ms}
	\right)-\nonumber \\
&&	-\frac{16\pi}{\omega^{\frac{n-2}{n}}}
	(-f)^{\frac{1}{n}}
	(N^m_{ma}f^{ab}N^n_{mb}-N^m_{ma}f^{ab}N^n_{nb}) \ . 
	\nonumber \\
\end{eqnarray}
If we combine these two equations together we get 
\begin{equation}
	\partial_c (N^c_{ab} f^{ab} ) = 0 \:
\end{equation}
that shows that  $\mG^c\approx 0$ is conserved during time evolution without any restriction on the value of Lagrange multiplier $\Lambda_c$. In other words $\mG^c\approx 0$ can be interpreted as the first class constraint. Then Lagrange multipliers $\Lambda_c$ will be determined by following way.
Firstly we contract the first equation in (\ref{eqmo}) with $f_{ab}$ and we obtain
\begin{eqnarray}
	f_{ab}\partial_c f^{ab}=\Lambda_c n \ , \nonumber \\
\end{eqnarray}
so that we can  express  $\Lambda_c$ as
\begin{equation}\label{Lambdac}
	\Lambda_c = 
	\frac{1}{n f}  \partial_c \det f \ . 
\end{equation}
 The situation
simplifies even more when we impose the condition
\begin{equation}
	\mF\equiv -\det f-K=0 \ , 
\end{equation}
where $K$ is constant. This constraint determines the value of the determinant of matrix $f^{ab}$ and
it is known as unimodular constraint.  Then from (\ref{Lambdac})  we immediately get $\Lambda_c=0$. 

On the other hand we should interpret $\mF$ as  gauge fixing constraint. Such a constraint has to be added  into action multiplied by appropriate  Lagrange multiplier $\Omega$ in order to be consistently included into dynamics. Since $\mF$ does not depend on $N^c_{ab}$ it is clear that the variation of $\mF$  only contributes to the equations of motion for $N_{ab}^c$ by factor $\Omega \frac{\delta \mF}{\delta f^{ab}}$. Explicitly, the equations of motion for $N_{ab}^c$ are modified by following way 
\begin{eqnarray}
	&&    0 = \partial_c N^c_{ab} + \Lambda_c N^c_{ab} + \nonumber \\
	&&    + \frac{16\pi}{n \omega^{\frac{n-2}{n}}} (-f)^{\frac{1}{n}} f_{ab} (N^m_{cd}f^{dn}N^c_{mn}-N^m_{ms}f^{sr}N^n_{nr})  +  \nonumber \\
	&&   +\frac{16\pi}{\omega^{\frac{n-2}{n}}}(-f)^{\frac{1}{n}} (N^m_{na} N^n_{mb} - N^m_{ma} N^n_{nb})  +\Omega f_{ab}(-f)
	\ .\nonumber \\
\end{eqnarray}
If we multiply this equation with $f^{ab}$ and use
the gauge fixing function $\mF$ we obtain 
\begin{eqnarray}
&&	0=\partial_c N^c_{ab}f^{ab}+\Lambda_c N^c_{ab}f^{ab}
	+ \frac{16\pi}{ \omega^{\frac{n-2}{n}}} (-f)^{\frac{1}{n}}  (N^m_{cd}f^{dn}N^c_{mn}-N^m_{ms}f^{sr}N^n_{nr})  +  \nonumber \\
&&	+\frac{16\pi}{\omega^{\frac{n-2}{n}}}(-f)^{\frac{1}{n}} (N^m_{na} f^{ab} N^n_{mb} - N^m_{ma}f^{ab} N^n_{nb})  +
	\Omega n (-f)
	\ .\nonumber \\
\end{eqnarray}
However if we combine this equation with 
\begin{equation}
	0=N^c_{ab} \partial_c f^{ab} -
	\Lambda_c N^c_{ab} f^{ab} -
	\frac{32\pi}{\omega^{\frac{n-2}{n}}}(-f)^{\frac{1}{n}}  
	\left(
	f^{db} N^a_{cd} N^c_{ab} - N^c_{cb} f^{bs} N^m_{ms}
	\right) \:,
\end{equation}
we get
\begin{eqnarray}
	0=\partial_c\mG^c+\Omega nK
\end{eqnarray}
so that the requirement of the preservation of the constraint $\mG^c\approx 0$
implies $\Omega=0$ .

 In summary, the gauge fixed equations of motion have the form 
\begin{eqnarray}
&&    0 = \partial_c N^c_{ab} +  
 \frac{16\pi}{n \omega^{\frac{n-2}{n}}} (-f)^{\frac{1}{n}} f_{ab} (N^x_{cd}f^{dy}N^c_{xy}-N^m_{ms}f^{sr}N^n_{nr})  +  \nonumber \\
&&   +\frac{16\pi}{\omega^{\frac{n-2}{n}}}(-f)^{\frac{1}{n}} (N^m_{na} N^n_{mb} - N^m_{ma} N^n_{nb}) \ , 
\nonumber \\
&&  \partial_c f^{ab}        -\frac{16\pi}{\omega^{\frac{n-2}{n}}}(-f)^{\frac{1}{n}}  
\left(
f^{db} N^a_{cd} + f^{da} N^b_{cd} - \delta^a_c f^{bs} N^m_{ms} - \delta^b_c f^{as} N^m_{ms}
\right)=0 \ . \nonumber \\
\end{eqnarray}
These equations of motion should correspond to the equations of motion of unimodular gravity that were derived recently in 
\cite{Kluson:2023yzv}. Unfortunately these two set of equations do not agree. In more details, we showed in \cite{Kluson:2023yzv} that consistency of the unimodular gravity in covariant formalism implies the presence of the secondary constraint $N^r_{ra}=0$ while in Weyl gravity there is a primary constraint $N_{ab}^c f^{ab}=0$. Further, the way how we determined Lagrange multiplier in \cite{Kluson:2023yzv}   is not exactly in the spirit of the analysis of the constraint systems due to the fact that in  the covariant canonical formalism it is not possible to solve equations $\partial_c N^d_{ab}$ since the equations of motion of covariant canonical formalism determine $\partial_c N^c_{ab}$ only. It is possible that the proper treatment of this problem could be in the powerful method developed by Kanatchikov in \cite{Kanatchikov:1993rp,Kanatchikov:2008he}. We hope to return to this analysis in near future. 

\subsection{Relation Between Surface and Bulk Lagrangians}
So far, we were concerned with the bulk part of Lagrangian. Now, we will focus on the surface part and find whether, there is a connexion between it and the bulk part. Such connexion can be found for Lanczos-Lovelock models as shown in \cite{Mukhopadhyay:2006vu}. In $F(R)-$Gravity, the connexion is not present \cite{Kluson:2022qxl}. Let us start with the boundary Lagrangian, which has the form of
\begin{equation}
\mL_{bound} = \frac{\omega}{16\pi} \partial_n \left(\tg^{ml} \tGamma^n_{ml}\right) \ . 
\end{equation}
We would like to connect it with the canonical momentum. From (\ref{Nab}) we find its contracted form as
\begin{equation}
    N^k_{ka} = \frac{\omega^\frac{n-2}{n}}{16\pi n (-f)^\frac{1}{n}} \tGamma^k_{rs} \tg^{rs} \tg_{ak} \: ,
\end{equation}
there is clearly visible the similarity between the surface Lagrangian and the contracted momentum. With little care one easily arrives to the relation
\begin{equation}
\mL_{bound} = \partial_b \left(
 n N^k_{ka} f^{ab}
\right) \:.
\end{equation}
We will discuss this relation in more details in the next section.
\section{Covariant Canonical Formalism for Weyl Gravity Formulated without Auxiliary Metric}\label{third}
In this section we develop covariant Hamiltonian formalism for Weyl gravity
that is formulated using the physical metric $g_{mn}$ instead of the metric $\tg_{mn}$.  
To do this we review basic facts about Weyl transformed metric in $n$ dimensions
\begin{equation}\label{trang}
 g'_{ij} = \Omega g_{ij}\  ,
 \end{equation}
where $\Omega$ is general function of space time. It is easy to see that under this transformation Ricci scalar $R'(g')$ is related to $R(g)$ through following formula 
\begin{eqnarray}\label{Rprime}
&&R'=\frac{1}{\Omega}R
+\frac{(1-n)}{\Omega}
\left(-\frac{1}{\Omega^2}\partial_i \Omega g^{ij}\partial_j\Omega+\frac{1}{\Omega}\frac{1}{\sqrt{-g}}
\partial_i[\sqrt{-g}g^{ij}\partial_j\Omega]\right)
\nonumber \\
&&+\frac{1}{4\Omega^3}
(n-2)(1-n)\partial_i \Omega g^{ij}\partial_j\Omega \ . 
\nonumber \\    
\end{eqnarray}
In case of Weyl gravity we have  $\Omega=(\frac{\omega^2}{-\det g})^{\frac{1}{n}}$
so that 
\begin{equation}
\partial_i \Omega=\frac{1}{n} \Omega \frac{\partial_i g}{-g} \ . 
\end{equation}
Then using (\ref{Rprime}) we get
\begin{eqnarray}
R'
=\frac{1}{\Omega}\left(R+\frac{(1-n)}{4n^2g^2}(5n-2)\partial_i g g^{ij}
\partial_j g+\frac{n-1}{ng}\frac{1}{\sqrt{-g}}\partial_i[\sqrt{-g}g^{ij}
\partial_j g]\right) \:, \nonumber \\
\end{eqnarray}
so that the action has the form 
\footnote{This action was analysed recently in \cite{Alonso-Serrano:2022rzj,Alvarez:2023zqz}.}
\begin{eqnarray}\label{Sweyltrans}
&&S=\frac{1}{16\pi \omega^{\frac{n-2}{n}}} \int d^nx (-g)^{\frac{1}{n}}
\left[R+\frac{(1-n)}{4n^2g^2}(5n-2)\partial_i g g^{ij}
\partial_j g+ \right.\nonumber \\
&&\left.+\frac{n-1}{ng}\frac{1}{\sqrt{-g}}\partial_i[\sqrt{-g}g^{ij}
\partial_j g]\right] \ .  \nonumber \\
\end{eqnarray}
Now we would like to express this action in the form that is suitable for covariant canonical formalism. First of all we use the fact that 
$\nabla_i g_{kl}=0$ that implies
\begin{eqnarray}
    \partial_i g_{kl}=\Gamma_{ik}^m g_{ml}+\Gamma_{il}^m g_{mk} \:,
\end{eqnarray}
that multiplied with $g^{kl}$ gives 
\begin{equation}
    \partial_i g=2\Gamma^k_{ik}g  \ . 
\end{equation}
Using this result we find
\begin{eqnarray}\label{secondpart}
 &&   (-g)^{\frac{1}{n}}
   \left(\frac{(1-n)}{4n^2g^2}(5n-2)\partial_i g g^{ij}\partial_j g
    +\frac{n-1}{ng}\frac{1}{\sqrt{-g}}\partial_i [\sqrt{-g}g^{ij}
    \partial_j g]\right)=\nonumber \\
\nonumber \\
&&=(-g)^{\frac{1}{n}}\frac{(1-n)(2-n)}{n^2}
\Gamma^m_{mi}g^{ij}\Gamma^n_{nj}+\frac{2(n-1)}{n}
\partial_i[(-g)^{\frac{1}{n}}g^{ij}\Gamma^k_{kj}] \ . 
\nonumber \\
\end{eqnarray}
Now we return to the first term in the action (\ref{Sweyltrans}) and perform the 
same manipulation as in previous section to obtain
\begin{eqnarray}\label{Rgn}
&&(-g)^{\frac{1}{n}}R=(-g)^{\frac{1}{n}}Q_k^{ \ mnl}R^k_{ \ mnl}=
2(-g)^{\frac{1}{n}}Q_k^{ \ mnl}[\partial_n \Gamma_{lm}^k+
\Gamma_{np}^k\Gamma^p_{lm}]=\nonumber \\
&&=(-g)^{\frac{1}{n}}
\left(\Gamma^m_{nk}g^{kl}\Gamma^n_{lm}-(1-\frac{2}{n})
\Gamma^n_{nk}g^{km}\Gamma^l_{lm}-
\frac{2}{n}\Gamma^m_{nk}g^{kn}\Gamma^l_{lm}\right)+\nonumber \\
&&
+2\partial_n[(-g)^{\frac{1}{n}}Q_k^{ \ mnl}\Gamma^k_{lm}] \ , 
\nonumber \\
\end{eqnarray}
where
\begin{eqnarray}
&&   R^k_{ \ mnl}=\partial_n \Gamma^k_{lm}-
\partial_l \Gamma^k_{nm}+\Gamma^k_{np}\Gamma^p_{lm}
-\Gamma^k_{lp}\Gamma^p_{mn} \ , \nonumber \\
&&Q_k^{ \ mnl}=\frac{1}{2}(g^{ml}\delta_k^n-g^{mn}\delta_k^l)
\ , \nonumber \\
\end{eqnarray}
and we used the fact that 
\begin{eqnarray}
    \partial_i (-g)^{\frac{1}{n}}=\frac{2}{n}\Gamma^{k}_{ki}(-g)^{\frac{1}{n}} \ .  \nonumber \\
\end{eqnarray}
If we then combine (\ref{secondpart}) with (\ref{Rgn}) we find that
the action (\ref{Sweyltrans}) can be written as 
\begin{eqnarray}
&&S=\frac{1}{16\pi}  \int d^nx 
(-g)^{\frac{1}{n}}
(\Gamma^m_{nk}g^{kl}\Gamma^n_{lm}+\frac{2-n}{n^2}
\Gamma^n_{nk}g^{km}\Gamma^l_{lm}-
\frac{2}{n}\Gamma^m_{nk}g^{kn}\Gamma^l_{lm})+
\nonumber \\
&&+\frac{1}{16\pi}\int d^nx\partial_n [(-g)^{\frac{1}{n}}
(g^{lm}\Gamma_{lm}^n+\frac{(n-2)}{n}g^{nm}\Gamma_{lm}^l)]\equiv \nonumber \\
&&\equiv 
\int d^nx (\mL_{bulk}+\mL_{bound}) \ ,  
\nonumber \\
\end{eqnarray}
where for simplicity we set $\omega=1$. 
Now we are ready to find conjugate momenta. We firstly define $M^{kmn}$ as 
\begin{eqnarray}\label{defM}
 &&   M^{kmn}=\frac{\partial \mL_{bulk}}{\partial(\partial_k g_{mn})}=
\frac{1}{16\pi}(-g)^{\frac{1}{n}}
\left[g^{mt}\Gamma_{tp}^k g^{pn}+
\frac{(2-n)}{n^2}g^{kp}\Gamma^s_{sp}g^{mn}-
\right.\nonumber \\
&&-\left.\frac{1}{n}\Gamma^k_{st}g^{st}
g^{mn}
-\frac{1}{n}(g^{kn}g^{mr}\Gamma_{pr}^p
+g^{nr}g^{mk}\Gamma^p_{pr}-
g^{kr}\Gamma^p_{pr}g^{mn})\right] \ .
\nonumber \\
\end{eqnarray}
Now using (\ref{defM}) 
we obtain
\begin{eqnarray}
    M^{kmn}g_{mn}=
 \frac{1}{16\pi}(-g)^{\frac{1}{n}}
 (g^{pt}\Gamma^k_{tp}+\frac{(2-n)}{n}
 g^{kp}\Gamma^s_{sp}-\nonumber \\
 -\Gamma^k_{st}g^{st}
 -\frac{1}{n}(2 g^{kn}g^{mr}
 \Gamma^p_{pr}g_{mn}-n g^{kr}\Gamma^p_{pr})=0 \:,
\end{eqnarray}
that implies an existence of primary constraints $M^{kmn}g_{mn}\approx 0$.
Then clearly it is possible to find covariant canonical  formulation of
 this theory with canonical variables
$g_{mn}$ and $M^{mn}$.  However we rather introduce variable $f^{ab}$ as in previous section where now we will be more general and consider following definition
\begin{equation}
f^{ab}=(-g)^\alpha g^{ab}  \ , 
\end{equation}
 where $\alpha$ is arbitrary number. 
 Then  conjugate momenta $N_{ab}^c$  are defined as
\begin{eqnarray}\label{defNabc}
&&    N^c_{ab}=\frac{\partial \mL_{quad}}{\partial (\partial_c f^{ab})}=
   \frac{\partial \mL_{quad}}
   {\partial (\partial_k g_{mn})}
   \frac{\partial (\partial_k g_{mn})}
   {\partial (\partial_c f^{ab})}=
   \nonumber \\
&&=M^{kmn}\delta_k^c\frac{\delta g_{mn}}{\delta f^{ab}}=
M^{kmn}\delta_k^c(-g_{mr}\frac{\delta g^{rs}}{\delta f^{ab}}
g_{sn})=\nonumber \\
&&=-M^{cmn}g_{mr}g_{ns}
(\frac{1}{2}(\delta^r_a\delta^s_b+\delta^r_b\delta^s_a)-
\frac{\alpha}{n\alpha-1}f^{rs}f_{ab})(-f)^{-\frac{\alpha}{n\alpha-1}}=\nonumber \\
&&=
-\frac{1}{16\pi}(-f)^{-\frac{1}{n}}
[\Gamma^c_{ab}-\frac{1}{n}\Gamma^c_{mn}f^{mn}f_{ab}
+\frac{2}{n^2}f^{cm}\Gamma^p_{pm}f_{ab}
-\frac{1}{n}(\delta_a^c\Gamma^p_{pb}+\delta^c_b\Gamma^p_{pa})] \:,
\nonumber \\
\end{eqnarray}
using the fact that 
\begin{equation}
(-g)=(-f)^{\frac{1}{n\alpha-1}} \ , 
    g^{ab}=f^{ab}(-f)^{-\frac{\alpha}{n\alpha-1}} \:,
\end{equation}
that also implies
\begin{equation}
\frac{\delta g^{rs}}{\delta f^{ab}}=
\left(\frac{1}{2}(\delta^r_a\delta^s_b+\delta^r_b\delta^s_a)-
\frac{\alpha}{n\alpha-1}f^{rs}f_{ab}\right)(-f)^{-\frac{\alpha}{n\alpha-1}} \ . 
\end{equation}
It is remarkable that the conjugate momentum $N^c_{ab}$ does not depend on $\alpha$. Further, from (\ref{defNabc}) we get 
\begin{eqnarray}
N^c_{ab}f^{ab}=-\frac{1}{16\pi}(-f)^{-\frac{1}{n}}
[\Gamma^c_{ab}f^{ab}-\frac{1}{n}
\Gamma^c_{mn}f^{mn}n+
\frac{2}{n^2}f^{cm}\Gamma^p_{pm}n-\frac{2}{n}
f^{cb}\Gamma^p_{pb}]=0 \:, \nonumber \\
\end{eqnarray}
that implies set of primary constraints
\begin{equation}
    \mG^c\equiv N^c_{ab}f^{ab}\approx 0 \ ,
\end{equation}
which are the same as in previous section.
Then the bare Hamiltonian is equal to
\begin{eqnarray}
&&\mH_B=\partial_c f^{ab}N^c_{ab}-\mL_{bulk}=
-\partial_c f^{ab}M^{cmn}g_{mr}\frac{\delta g^{rs}}{\delta f^{ab}}g_{ns}-\mL_{bulk}=
\nonumber \\
&&=\partial_kg_{mn}M^{kmn}-\mL_{bulk}=(\Gamma^p_{km}g_{pn}+\Gamma^p_{kn}g_{nm})
M^{kmn}-\mL_{bulk}=\nonumber \\
&&=\frac{1}{16\pi}(-g)^{\frac{1}{n}}
(\Gamma^m_{nk}g^{kl}\Gamma^n_{lm}+\frac{2-n}{n^2}
\Gamma^n_{nk}g^{km}\Gamma^l_{lm}-
\frac{2}{n}\Gamma^m_{nk}g^{kn}\Gamma^l_{lm}) \ . 
\nonumber \\
\end{eqnarray}
In order to find Hamiltonian as function of canonical variables we again calculate
\begin{eqnarray}
    N^c_{ab}g^{ad}N^b_{cd}=\bA^2
[\Gamma^c_{ab}g^{ad}\Gamma^b_{c}+\frac{-2n^2+2n-4}{n^3}\Gamma^m_{mr}\Gamma^r_{ts}g^{gs}+\nonumber \\
+\frac{(3n^2-n^3-4n+4)}{n^4}\Gamma^m_{ma}g^{ab}\Gamma^n_{nb}+\frac{1}{n^2}\Gamma^c_{mn}g^{mn}g_{cb}\Gamma^b_{pq}g^{pq}] \ , 
\nonumber \\
\end{eqnarray}
where $\bA=-\frac{1}{16\pi}(-f)^{-\frac{1}{n}}$.
We further have
\begin{eqnarray}
 &&   N^r_{ra}g^{ab}N^t_{tb}
 =\nonumber \\
 &&=\bA^2[\frac{(2-n)^2}{n^4}\Gamma^p_{pa}g^{ab}\Gamma^s_{sb}-2\frac{(2-n)}{n^3}\Gamma^p_{pa}\Gamma^a_{rs}g^{rs}+
 \frac{1}{n^2}\Gamma^a_{st}g^{st}g_{ab}\Gamma^b_{mn}g^{mn}] \ . 
 \nonumber \\
\end{eqnarray}
Collecting these terms together we find final form of the bare Hamiltonian 
\begin{eqnarray}
    \mH_B=16\pi(-f)^{\frac{1}{n}}
[N^c_{ab}f^{ad}N^b_{cd}-N^r_{ra}f^{ab}N^t_{tb} ] \ ,\nonumber \\
\end{eqnarray}
which has exactly the same form as the Hamiltonian density derived in previous section. We would like however stress one important point which is the fact that we used 
generalized form of the variable $f^{ab}=(-g)^\alpha g^{ab}$ and 
that the theory does not depend on $\alpha$ at all.

Finally we return to the boundary term that is equal to
\begin{equation}
\mL_{bound}=\frac{1}{16\pi}\partial_n [(-g)^{\frac{1}{n}}
(g^{lm}\Gamma_{lm}^n+\frac{(n-2)}{n}g^{nm}\Gamma_{lm}^l)] \ . 
\end{equation}
Since $N^r_{ra}$ is equal to
\begin{equation}
N^r_{ra}=\frac{1}{16\pi}(-f)^{-\frac{1}{n}}
[\frac{(n-2)}{n^2}\Gamma^p_{pa}+\frac{1}{n}
\Gamma^c_{mn}f^{mn}f_{ca}] \:,
\end{equation}
we again find 
%
that the surface term has the form 
\begin{equation}
\mL_{bound}=n\partial_n[f^{nm}N_{rm}^r] \:,
\end{equation}
that agrees with the result derived in previous section. Further, 
 this expression can be written as
\begin{eqnarray}
    \mL_{bound}=n\partial_n\left[f^{nm}\delta_c^r
    \frac{\partial \mL_{bulk}}{\partial (\partial_c f^{rm})}\right] \:, \nonumber \\
\end{eqnarray}
that has the form of holographic relation between bulk and boundary action with 
agreement with the general discussion presented in  
\cite{Mukhopadhyay:2006vu}.

{\bf Acknowledgement:}

The work of JK is supported by the grant “Dualitites and higher order derivatives” (GA23-06498S) from the Czech Science Foundation (GACR).


\end{document}